# Optical-parametric-amplification-enhanced background-free spectroscopy

**Mingchen Liu[1], Robert M. Gray[1], Arkadev Roy[1], Luis Ledezma[1], and Alireza Marandi[1, *]**

[1]*Department of Electrical Engineering, California Institute of Technology, Pasadena, California, 91125, USA*
[*]*marandi@caltech.edu*

**Abstract:** Traditional absorption spectroscopy has fundamental difficulty in resolving small absorbance from strong background due to the instability of laser sources. Existing background-free methods in broadband vibrational spectroscopy help to alleviate this problem but face challenges in realizing either low extinction ratios or time-resolved field measurements. Here, we introduce optical-parametric-amplification-enhanced background-free spectroscopy, in which the excitation background is first suppressed by an interferometer and then the free-induction decay that carries molecular signatures is selectively amplified. We show that this method can further improve the limit of detection in linear interferometry by order(s) of magnitude without requiring lower extinction ratios or time-resolved measurement, which can benefit sensing applications in detecting trace species.

Optical absorption spectroscopy is a powerful and versatile tool to study properties of different materials. The absorption information is typically contained in the change of the radiation source and deciphered by comparison between at least two measurements of the optical spectrum. Detecting trace samples with low concentrations is important and can push the limits of a wide range of applications, such as breath analysis [1], industrial control [2] and environmental monitoring [3]. However, in traditional absorption spectroscopy, detection of tiny absorption dips on top of a large background is a fundamental challenge, which is limited by the noise and stability of the light source, as well as the dynamic range of the whole detection system.

There are some existing background-free spectroscopy (BFS) methods, including photoacoustic spectroscopy [4,5], Faraday rotation spectroscopy [6,7], and laser-induced fluorescence spectroscopy [8]. Nevertheless, they are limited either in access to narrow resonances and quantitative measurement capabilities or only applicable to a small class of molecules and narrow wavelength range. These challenges have limited such techniques to prototypical demonstrations in laboratory settings in contrast to more standard infrared spectroscopy techniques like Fourier transform infrared spectroscopy (FTIR).

Recently, to realize a background-free detection in broadband infrared (ro-vibrational) spectroscopy, two types of approaches have been proposed and demonstrated. The first is temporal gating based on short excitation pulses and nonlinear wave mixing [9–13], in which the excitation background is detected and separated from the free-induction decay signal directly in the time domain. However, time-resolved measurements require not only accurate synchronization (femto- or even atto-second level) and scanning between two independent pulse trains but also super short pulses, which may have to be shorter than one optical cycle of the excitation pulse. These components are challenging to realize and necessitate substantial efforts.

The second is broadband linear interferometry [14–16], which is motivated by LIGO [17], dual-beam interferometry [18] and some narrowband laser absorption spectroscopy works [19,20]. In this approach, a Mach-Zehnder-like or Michelson-like interferometer arranged for destructive interference is used to coherently subtract the background from the optical field using a sign-inverted replica before the optical power arrives at the photodetector, which converts absorption from dips to peaks in spectra. However, this method is directly limited by the realistic intensity extinction ratio (field unbalanced factor), which necessitates locking and additional components in the setup to control and practically difficult to further decrease. Therefore, the advantage of this BFS method over direct absorption spectroscopy (DAS) is limited to only a ~10 times improvement in SNR [14,16] and is not experimentally demonstrable in some cases [15].

In this work, we propose a new method, optical-parametric-amplification-enhanced background-free spectroscopy (OPA-BFS). We discuss it in the context of ro-vibrational spectroscopy, but it is also potentially applicable to other kinds of absorption spectroscopy. First, similar to refs. [14–16], the sample is interrogated by short pulses (generally mid-IR), the background excitation of which is suppressed by an interferometer. Next, the output from the interferometer, which includes sample response and residual background, is amplified by a short-pulse optical parametric amplifier (OPA). The pump pulses (generally near-IR) of the OPA are kept at a chosen delay relative to the signal pulses (output from the interferometer), so they can amplify a strong part of the FID field while being far away from the center of the original excitation pulses to avoid residual background. We theoretically and numerically demonstrate that this method can further improve the SNR and limit of detection (LOD) of the above-mentioned broadband linear BFS by orders of magnitude, without requiring a lower extinction ratio or time-resolved measurements which can be experimentally challenging. On one hand, while OPA-BFS amplifies the absorption signal of the samples and make it more detectable, there is no limitation on the type of spectrometer used for spectrum acquisition; one can either use a typical frequency-domain spectrometer, like a grating-based OSA, monochromator, or FTIR, or a time-domain spectrometer, such as dual-comb spectroscopy [21], electro-optic sampling [12,22] or cross-comb spectroscopy [13]. In comparison, existing BFS by temporal gating [9,11–13] is less flexible because it demands time-resolved spectrometry, which can have some advantages over traditional frequency-domain spectrometry but requires more experimental effort. On the other hand, thanks to the temporal gating provided by short-pulse nonlinearity, OPA-BFS is not limited by and has a relaxed requirement on the extinction ratio of the destructive interference compared to existing broadband BFS based on linear interferometry. Although extinction ratios of ~$10^{-4}$ have been demonstrated [14–16], achieving further extinction remains technically challenging due to misalignment, substrate thickness mismatch and environment noise [15], which strictly limits the advantages of linear BFS.

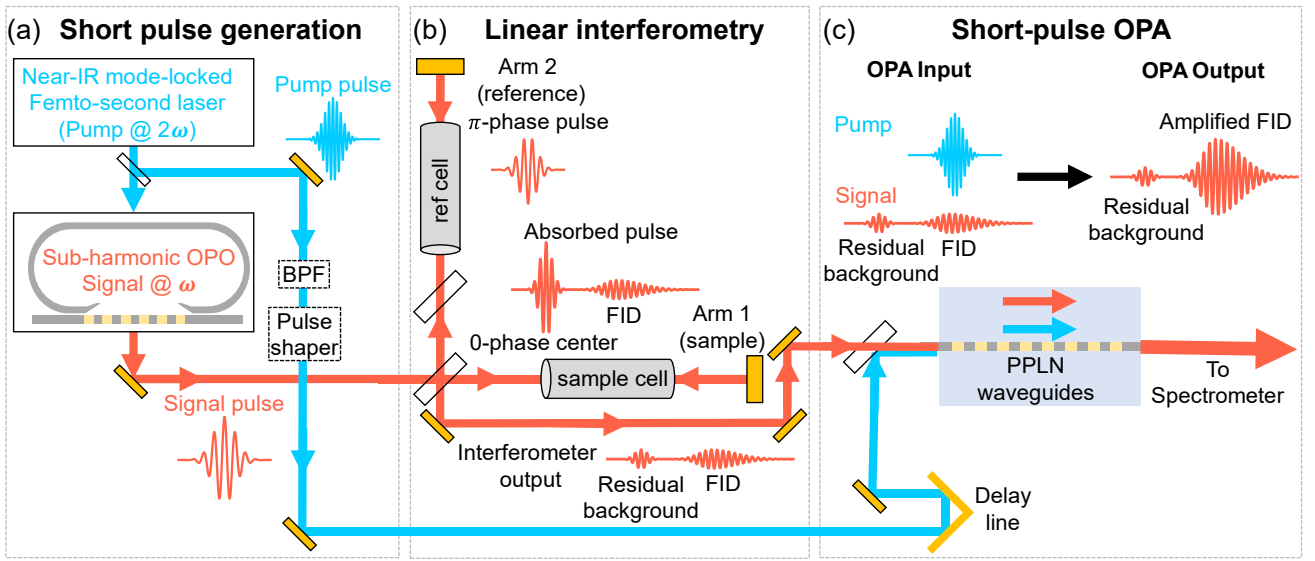

**Fig. 1.** Optical-parametric-amplification-enhanced background-free spectroscopy (OPA-BFS). (a) short pulse generation. BPF: bandpass filter. BPF and pulse shaper may be required to change and control the profile and pulse width of the original pump pulse because the short-pulse OPA in (c) may need a pump pulse with a longer pulse width and a different profile. (b) Linear interferometry. While a Michelson-like interferometer is illustrated here, a Mach-Zehnder-like interferometer can also work. For clarity, we only present the most important components of the interferometer; more details, especially regarding dispersion compensation and delay locking, can be found in Ref [14–16]. Note that we make a very short and clean separation between the excitation pulse (center) and FID radiation for clarity of the illustration, which is not always the case in practice. However, this will not influence our following analysis and arguments, as there will always be part of the FID radiation that is far enough from the excitation pulse center and thus can be separated well. (c) Short-pulse OPA. Here, we show the illustration of an OPA based on nanophotonic periodically-poled lithium niobate (PPLN) waveguides [23,24], which was recently demonstrated with unprecedented high gain and broad bandwidth. However, it can also be any other platform or material that can support short-pulse OPA with high parametric gain.

The architecture of OPA-BFS is presented in Fig. 1, which is composed of three parts: short pulse generation, linear interferometry, and short-pulse OPA. While OPA-BFS does not require any specific technique for the pulse generation, Fig. 1(a) illustrates a sub-harmonic optical parametric oscillator (OPO) synchronously pumped by a short-pulse mode-locked laser (typically a fiber laser), which is a common way to generate short mid-IR pulses [25–29]. One important advantage of synchronously-pumped OPOs is that the timing and phase of signal pulses and pump pulses are intrinsically locked, which can exempt additional efforts in their control for the short-pulse OPA [30]. The second step is to use the signal pulses (generally mid-IR) to interrogate the sample with a detection background suppressed by linear interferometry, as illustrated in Fig. 1(b). The output of the interferometer consists of two parts, the residual pulse center (background) which cannot be fully eliminated by the interferometer and the subsequent FID signal which carries the spectral information of the sample. Compared to the residual background (originally the excitation pulse) that is much more localized in the time domain (pulse width of ~10-100 fs), the FID signal can typically last at least hundreds of ps and sometimes have a local maxima at a relative delay of 10-100 ps [9,10,31,32]. This can be understood equivalently in the frequency domain; while femtosecond pulses can have a bandwidth as broad as tens of THz, a typical vibrational absorption has a linewidth on the order of magnitude of only 10 GHz at room temperature and atmospheric pressure, which can be even smaller at lower pressure or temperature. The output of the interferometer is then sent to a short-pulse OPA (Fig. 1(c)) as the signal to be amplified. The pump pulse is held at a chosen delay with respect to the signal such that it overlaps with a strong portion of the FID but is far away from the excitation center. Therefore, the FID carrying useful sample signatures is amplified while the residual background is not, as it does not temporally overlap with the pump pulse. This can further improve the SNR of the absorption spectrum and make a trace sample detectable that cannot be detected by DAS or linear BFS.

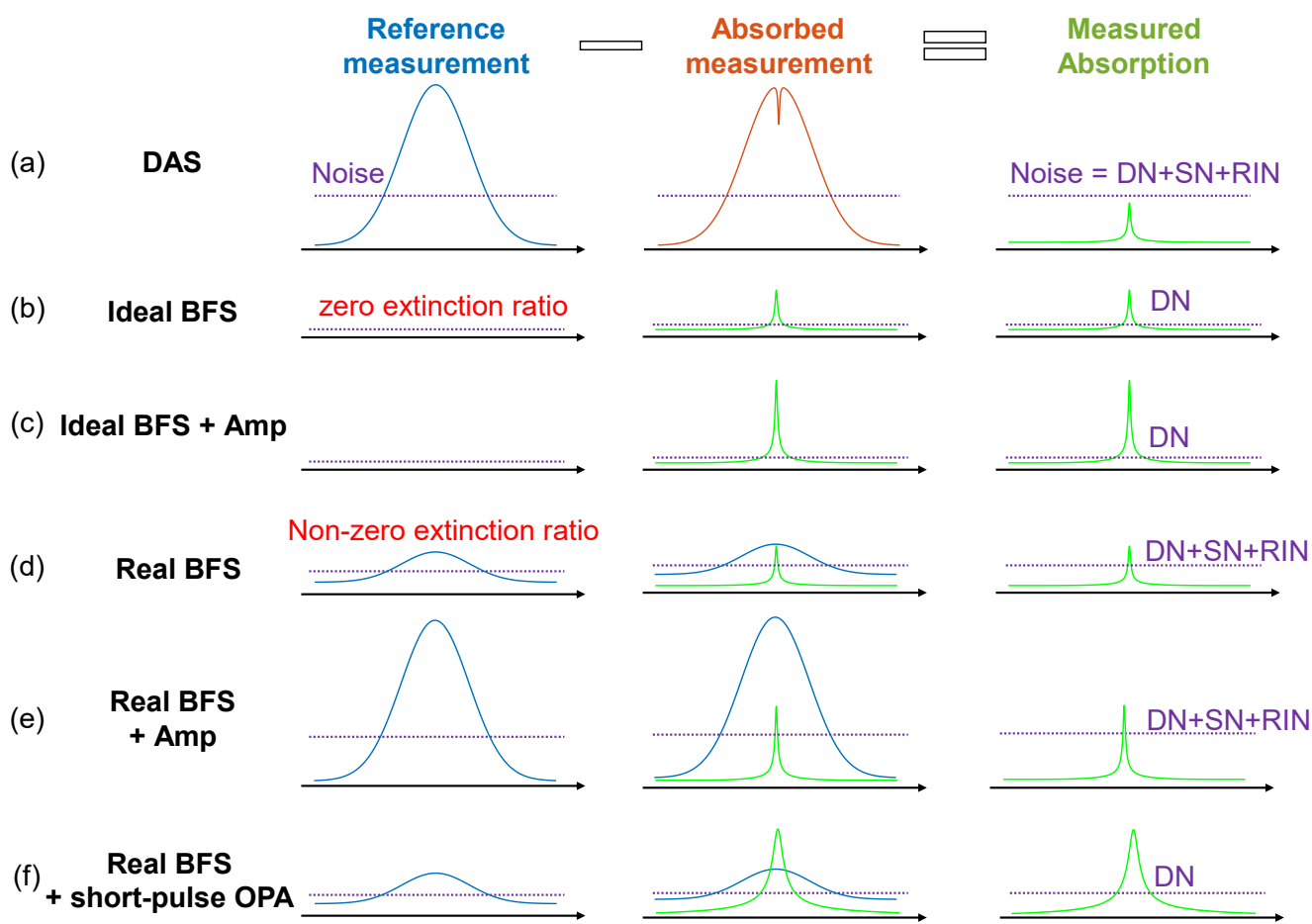

Fig. 2. Qualitative comparison between different spectroscopy schemes. (a) DAS. (b) Ideal (linear) BFS. (c) Ideal BFS followed by an ideal general frequency-domain amplifier (GA). (d) Real BFS. (e) Real BFS followed by a GA. (f) Real BFS followed by a short-pulse OPA.

Figure 2 qualitatively compares different spectroscopy schemes in detecting small absorption (trace sample) to show the advantage of OPA-BFS. In traditional DAS (Fig. 2 (a)), one must compare two measurements, one without sample (reference measurement, blue curve) and one with sample (absorbed measurement, red curve), the difference of which is the absorption signal of interest (green curve). There are three primary kinds of noise, detector noise (DN), shot noise (SN) and relative intensity noise (RIN), and any of them may dominate and limit the detection depending on the power incident on the detector. The noise level for each spectrum is denoted by the purple dashed line. Generally, if we assume a high source power which can saturate the detector, the RIN will dominate and fundamentally limit the detection. Therefore, one cannot detect an absorption dip smaller than the RIN, which is proportional to the full power of the light source (excitation background). In ideal BFS (Fig. 2(b)), the excitation background can be fully eliminated in the reference measurement, and the absorption is converted from dip to a peak in the absorbed measurement. In this case, the absorption peak needs to overcome just the DN to be detectable, which is the only noise that may limit the detection. Note that this does not mean RIN and SN do not exist in the measurement; they are always present and proportional to the power (RIN) or square of the power (SN) incident on the detector. However, in this case, all power arriving at the detector is from the absorption, the signal of interest, so a noise which is proportional to the signal such as SN or RIN cannot limit the detection. Therefore, DN is the only possible limiting factor, which is why we only indicate DN there (same for Figs. 2(c) and (f)). As absorption is a peak instead of a dip in BFS, it is natural to add an amplifier after it, which can further improve the detectability of the signal (Fig. 2(c)). Note that the amplifier here refers to a general ideal frequency-domain amplifier (GA) which does not bring in extra noise and has no temporal features. In these two ideal cases, the LOD (defined as minimum detectable absorbance) is free from detector saturation and RIN or SN and purely decided by the available source power and amplification (if applicable).

However, these two ideal cases are not realistic because the extinction ratio in real linear interferometry is always non-zero and results in a residual background. Although linear BFS can increase the SNR to some extent (Fig. 2(d)), the residual background can still fundamentally limit the LOD via SN or RIN at high power like DAS and prevents detection of lower absorption. Moreover, adding a general (frequency-domain) amplifier after the linear interferometry is not helpful in the RIN-limited regime because the noise (SN and RIN) from the residual background is also amplified by the same factor as the absorption signal (Fig. 2(e)). In contrast, a short-pulse OPA can make a difference (Fig. 2(f)). Upon a proper timing of the pump pulse, one can amplify only the FID (absorption signal) but avoid the excitation pulse center (residual background), which will remain almost the same. Thus, OPA-BFS can further increase the SNR in addition to the enhancement from linear BFS, and the LOD of OPA-BFS is fundamentally limited by the available source power and OPA amplification.

To further demonstrate the advantages of OPA-BFS quantitatively, we conduct theoretical analysis and numerical simulation for different detection schemes and types of samples. The mathematical models for BFS of linear interferometry and noise analysis are based on ref [14–16,33]. Specifically, some important parameters are adapted from a recent state-of-the-art experimental result reported in ref [16], including the field unbalanced factor $\delta = 10^{-2}$ and RIN ratio $\sigma_r = 10^{-2}$. Note that those numbers are very close to the experimental results in ref [16] but with a simpler value for ease of presentation. More importantly, for linear BFS, the model using those two parameters gives a theoretical LOD of absorbance equal to $\delta\sigma_r = 10^{-4}$, which agrees with what is experimentally demonstrated in ref [16]. The simulation of OPA is based on solution of the coupled wave equations using the Fourier split step method [25,34], the parameters of which are based on previous experimental demonstrations of high-gain OPA in thin-film lithium niobate [23]. The absorption of molecules is modeled based on data from the HITRAN database [35], using a Lorentz oscillator model for the line profile. Detailed description and parameters of our theoretical and numerical models can be found in the Supplementary Information.

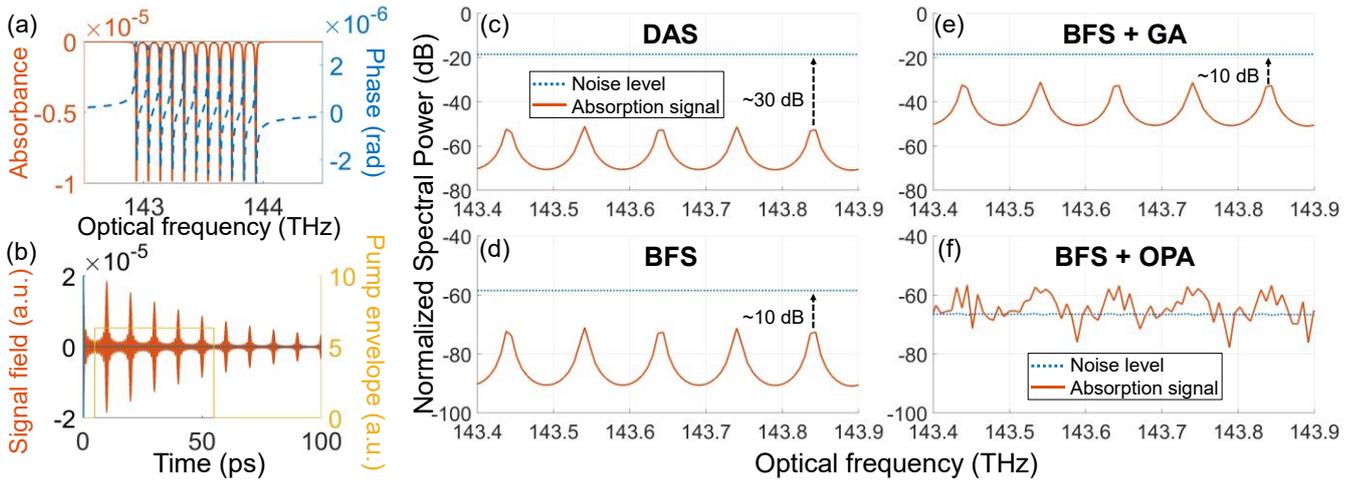

**Fig. 3.** OPA-BFS for a mock sample. (a) Intensity (absorbance, red solid curve) and phase (blue dashed curve) of 11 Lorentzian transitions assumed for the mock sample. (b) FID field of the signal pulse (red curve, left y-axis) that probed the sample in the interferometer (see "Arm 1" in Fig. 1(b)). Note that, to show the weak and long FID, the y scale (intensity) is zoomed in and x scale (time) is zoomed out; therefore the stronger and narrower background residual (blue curve) cannot be seen clearly here. The yellow curve (right y-axis) denotes the envelope of the pump pulse. (c)-(f) Spectral noise level (blue dotted curves) and ideal absorption signal (red solid curves) in different detection schemes. The absorption signal is the difference between the reference measurement (without sample) and the absorbed measurement (with sample), and the noise level is an incoherent addition (quadratic mean) of the total noise level (including DN, SN, and RIN) in these two measurements. Note that we zoom into the central five transitions to show the details more clearly.

First, we conduct a simulation for a mock sample to give a simple and clear illustration. The mock sample is set to have 11 equally strong and equally spaced Lorentzian transitions with the same linewidth of 6 GHz. Those transitions are distributed from 143 THz to 144 THz, with a peak absorbance of $10^{-5}$ as shown in Fig. 3(a), together with the phase profile. We use a sech signal (excitation) pulse with a center wavelength of 2.09 μm (143.4 THz) and a 40-fs pulse width to interact with the sample in the interferometer. The output of the interferometer consists of two parts, the residual background (excitation pulse) and FID, part of which is shown in Fig. 3(b) (red curve, left y-axis). One can observe a pattern in the FID with a period of 10 ps, which is a result of the coherent addition of reradiation of those transitions with a 100-GHz spacing [9,31,32]. To amplify the FID, we use a rectangular pump pulse with a center wavelength of 1.045 μm (286.9 THz) and pulse width of 50 ps, the envelope of which is denoted by the yellow curve (right y-axis) in Fig. 3(b). Note that we keep the center of the rectangular pump pulse at a delay of 30 ps with respect to the center of the signal pulse (zero of the time axis), by which the pump can cover a strong part of the FID while avoiding the residual background.

While a more detailed description about the simulation can be found in the Supplementary Information, the results of the absorption signal and noise level in different detection schemes are presented in Figs. 3(c)-(f). Note that we assume a grating-based spectrometer for detection of the 2.09-μm signal spectra with a resolution of 0.1 nm. Also, we assume a high enough average power for the 2.09-μm excitation pulse such that the peak of its spectrum can just saturate the detector of the spectrometer, and all the spectral power is normalized to it and presented on a logarithmic scale. Therefore, the noise level is about -20 dB in DAS (Fig. 3(c)), which is dominated by RIN, corresponding to a $\sigma_r = 10^{-2}$. As we set an absorbance of $10^{-5}$, the absorption signal level is -50 dB, which is 30-dB weaker than the noise level and thus undetectable in DAS. In linear BFS, while the absorption signal will be lowered by $\delta$, the background will be suppressed by $\delta^2$ as will the RIN (See Supplementary Information for detailed theoretical derivation). Therefore, the SNR can be increased by $1/\delta$ if the RIN still dominates, which is the case for our example here. As shown in Fig. 3(d)), compared to DAS, the noise level in BFS is suppressed by 40 dB ($\delta^2 = 10^{-4}$) and now around -60 dB, and the signal level is lowered by 20 dB ($\delta = 10^{-2}$) and now -70 dB. Obviously, although the SNR has been increased by 20 dB ($1/\delta$), the signal is still below the noise level and thus still undetectable. This agrees with the fact that the absorbance we set here ($10^{-5}$) is lower than the LOD of the linear BFS ($\delta\sigma_r = 10^{-4}$). Next, we try to amplify the output of the interferometer with an ideal general amplifier (GA) with a power gain of 40 dB, and the result is shown in Fig. 3(e). Compared to BFS, while the signal is amplified by 40 dB, the noise level is also amplified by the same factor; therefore, the SNR is not increased. We assume a power gain of 40 dB because it corresponds to $(1/\delta)^2$, which will bring the output spectra back to the saturation level of the spectrometer. One can apply a higher power gain if the spectrometer saturation is not considered, but it still cannot increase the SNR. Finally, Fig. 3(f) presents the result of OPA-BFS. The absorption signal reaches above -60 dB, which is amplified by about 10 dB in the frequency domain. This frequency-domain amplification factor is much less than that of the GA in panel (e) because the OPA pump pulse only covers a small temporal range of the whole FID. However, as the pump pulse avoids the residual background in the time domain, the noise is not amplified like GA, by which the SNR is effectively increased compared to linear BFS. In fact, we even observe a decrease in the noise, because part of the energy of the residual background pulse (2.09 µm) flows to the pump wavelength (1.045 µm) via second harmonic generation (SHG). The SHG here is prominent since the signal pulse is set with a relatively high power because we want to work in the RIN-limited regime. Even if we ignore the SHG effect, the absorption signal reaches above -60 dB, the same as the noise in the linear BFS (see Fig. 3(d)), and so will still be detectable (SNR >= 1). In this case, the LOD will be limited by the amplification of the OPA instead of the RIN or detector saturation. There is some observed broadening and distortion of the resolved peaks, which is mainly due to the finite temporal window of the pump pulse and the phase-sensitive nature of the OPA gain. Nevertheless, the basic spectral information, including the center frequencies and relative intensities of the transitions, are well preserved. One can always try to apply a longer pump pulse to cover a wider temporal range to alleviate this problem. However, for a given average power of the pump, there is a trade-off between the peak power (temporal gain) and width (temporal window) of the pump pulse.

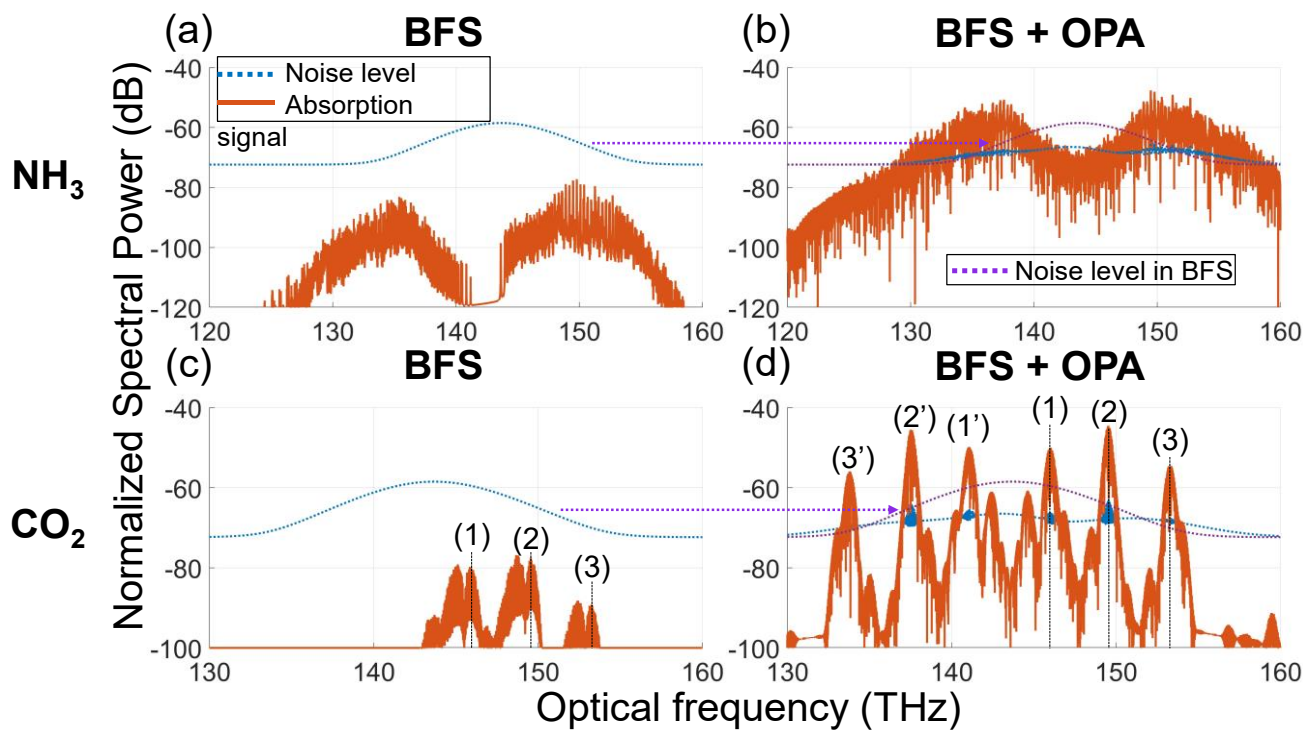

**Fig. 4.** BFS and OPA-BFS for $NH_3$ ((a)-(b)) and $CO_2$ ((c)-(d)) around 143.4 THz. The purple dotted lines in (b) and (d) denote the noise level in corresponding BFS spectra (blue dotted curves in (a) and (c)). Note that there are some weak transitions missing around the center of the $NH_3$ absorption (141-142 THz) due to data missing from HITRAN database.

For further demonstration of OPA-BFS, we conduct simulation with real molecules. As we have shown that the SNR of BFS can be higher than that of DAS and cannot be further increased by an ideal GA, here we only present the results of BFS and OPA-BFS. Note that the parameters for the linear BFS and noise are the same as those of the last example. Results for $NH_3$ are shown in Figs. 4(a)-(b). We set the transition at 151.3 THz, the strongest one around the center frequency of our excitation pulse (143.4 THz), to have an absorbance of $10^{-5}$, so the absorbance of other nearby transitions is smaller than $10^{-5}$. Therefore, all transitions are below the LOD of the linear BFS (see Fig. 4(a)). Here, we continue to use a rectangular pump pulse but with a shorter pulse width of 20 ps, the center of which is held at a delay of 12 ps. As shown in Fig. 4(b), the absorption signal is above the noise level and well detectable in OPA-BFS. As before, the noise level in OPA-BFS (blue dashed curve) is decreased because of SHG. However, the absorption signal here is also higher than the original noise level in BFS (purple dashed line) and thus still detectable even if we do not consider the SHG effect. The same is observed in the case of $CO_2$ (see Fig. 4(c)-(d)). For $CO_2$, there are three groups of transitions of close to 143.4 THz, which are around 145.8 THz, 149.6 THz, and 153.3 THz, as labeled in Fig. 4(c). The transition at 149.6 THz, the strongest among the three groups, is set to have an absorbance of $10^{-5}$. Here, different from previous cases, we use a sech pump pulse with a pulse width of 5 ps at a relative delay of 25 ps to amplify the FID. Fig. 4(d) shows that OPA-BFS makes the absorption signal stronger than the noise and thus readily detectable. Although there are some distortions to the absorption profile as we use a relatively short pump pulse for higher gain, the center of each transition group is well captured (see the labels corresponding to panel (c)), which can be enough for detection and identification of the molecule. Notably, new frequency components (labeled (1'), (2'), and (3')) are found on the other half of the spectrum. They are idler radiation generated in the OPA and thus symmetric to their corresponding signal frequencies with respect to the center frequency (143.4 THz). If we also include the radiation around the idler frequencies into our detection, the SNR and LOD can be further enhanced. In short, for both molecules, we demonstrate that OPA can enhance the LOD of linear BFS by more than one order of magnitude, considering that the absorbance for both molecules is less than or equal to $10^{-5}$.

Note that the obtained absorption signal in OPA-BFS depends on many parameters, including the power, profile, width, and center delay of the pump pulse, and we only show one possibility for each example above. Moreover, there is a trade-off between temporal gain and spectral resolution, the essence of which is the trade-off between width and peak power of pump pulse with a fixed average power. A complete and systematic discussion and optimization of those parameters is useful but involved, so they are beyond the scope of this work and will be the subject of future works.

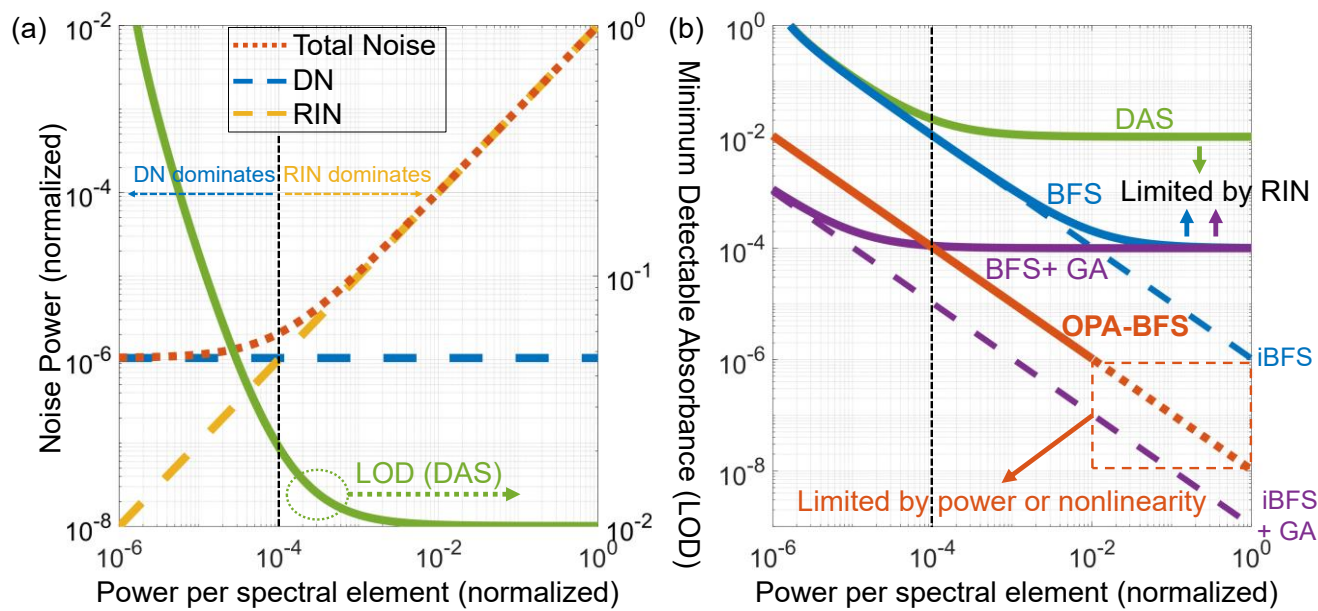

**Fig. 5.** Noise and LOD scaling with excitation power of different spectroscopy schemes. (a) Detector noise (DN, blue dashed line), relative intensity noise (RIN, yellow dashed line) and total noise (red dotted curve) in DAS (left y-axis). Green curve (right y-axis) denotes the limit of detection (minimum detectable absorbance, SNR=1) of DAS. (b) LOD scaling with excitation power of different schemes. Solid curves: DAS (green, same as the green solid curve in (a)), BFS (blue), BFS+GA (purple), and OPA-BFS (red). Dashed curves: ideal BFS (iBFS, blue) and iBFS+GA (purple).

We finally investigate how the noise and LOD scale with excitation power for different detection schemes, which are presented in Fig. 5. Here, we assume a detector saturation power of 0.1 mW and all power displayed is normalized to it. More details and parameters for this calculation can be found in the Supplementary Information. Figure 5(a) first depicts the scaling of the noise (left y-axis) for DAS as an example. In DAS, the total noise is dominated by detector noise (DN) or RIN when the relative excitation power is smaller or larger than $10^{-4}$, respectively. This power scaling is basically similar to Fig. 1 of ref. [33] despite two differences. One is that we do not consider dynamic range of the whole detection system. The other is that the shot noise in our case is negligible and thus not shown in the figure, which is consistent with the finding in ref [16]. If we define "detectable" as SNR=1, the corresponding LOD for DAS can be calculated and is denoted by the green solid curve (right y-axis). While higher power leads to a lower LOD when the DN dominates, the LOD stops decreasing and converges to $10^{-2}$ ($\sigma_r$) as the RIN dominates. Following DAS, the LOD scaling of other schemes is depicted in Fig. 5(b). The LOD of BFS (blue solid curve) can be lower than that of DAS because of the RIN suppression, but it is still ultimately limited by RIN and converges to a fixed lower bound of $10^{-4}$ ($\sigma_r \delta$). An ideal general amplifier can decrease LOD of BFS at low power (purple solid curve), but it stops helping at higher power as the detector becomes saturated and the detection is limited by the RIN in the same way as the case without amplification. Finally, the red solid curve denotes the LOD of OPA-BFS. At low power, OPA is not as helpful as GA due to a gain penalty we set with it, since short-pulse OPA amplifies only a part of the signal in the time domain. However, OPA-BFS outperforms GA-BFS and linear BFS at higher power as its LOD continues to scale down because it is not limited by RIN. When the excitation power per spectral element is higher than a specific limit, which would vary case by case and we set $10^{-2}$ in this figure, further scaling down of the LOD in OPA-BFS (dotted red curve) depends on the availability of the total excitation pulse power or on having enough parametric gain for a high-power signal input. Note that two dashed lines, blue for ideal BFS (iBFS) and purple for iBFS+GA, are also displayed as useful references although they are not practical. In short, although BFS can lower the LOD of DAS to some extent, it is still limited by RIN at high powers due to a non-zero extinction ratio. While a general amplifier cannot effectively help, a short-pulse OPA can further lower the LOD of BFS by order(s) of magnitude.

The above discussion shows how short-pulse OPA-BFS can be practically useful compared to GA-BFS due to its ability to selectively amplify a portion of the time-domain signal, resulting in temporal gating. This motivates consideration of the method in comparison to other nonlinear sensing techniques, which may also provide temporal gating or up-conversion capabilities. One distinct advantage of OPA is its unique ability to achieve exponential amplification in the signal due to the conversion of photons from the pump [36,37], with amplification factors on the order of 100 dB/cm having been readily achieved [23]. This makes its gain and efficiency much higher than techniques based on, for example, second-harmonic generation or sum-frequency generation for the measurement of ultraweak signals [12,13,22], for which the output photon number cannot exceed the input signal photon number, placing a fundamental limit on the potential amplification [38]. OPA-BFS may also be considered in the non-degenerate regime, where signal up- or down-conversion is possible in addition to amplification and background can be intrinsically zero even without linear interferometry.

In summary, we introduce a new method named OPA-BFS. While it can achieve a higher SNR and lower LOD in broadband vibrational spectroscopy, it does not require a lower extinction ratio or time-resolved measurements, which is experimentally challenging but has remained essential to existing BFS works. OPA-BFS not only combines and improves upon many merits of demonstrated techniques for background-free vibration spectroscopy, including both linear and nonlinear ones, but also circumvents some of their practical challenges. This work sheds new light on the potential for detection of trace molecules enhanced by optical nonlinearity, which can enable new limits in broadband vibrational spectroscopy and benefit numerous applications. Recently, there have been substantial advances in high-power and broadband mid-IR femtosecond pulse generation [25,27,28] and unprecedented optical nonlinearity enabled by lithium niobate nanophotonics [23,24,39], which can enable experimental realization of this technique on both free-space and on-chip platforms.

**Reference**

1. Q. Liang, Y.-C. Chan, P. B. Changala, D. J. Nesbitt, J. Ye, and J. Toscano, "Ultrasensitive multispecies spectroscopic breath analysis for real-time health monitoring and diagnostics," Proceedings of the National Academy of Sciences **118**, e2105063118 (2021).
2. M. Jamrógiewicz, "Application of the near-infrared spectroscopy in the pharmaceutical technology," Journal of Pharmaceutical and Biomedical Analysis **66**, 1–10 (2012).
3. B. Nozière, M. Kalberer, M. Claeys, J. Allan, B. D'Anna, S. Decesari, E. Finessi, M. Glasius, I. Grgić, J. F. Hamilton, T. Hoffmann, Y. Iinuma, M. Jaoui, A. Kahnt, C. J. Kampf, I. Kourtchev, W. Maenhaut, N. Marsden, S. Saarikoski, J. Schnelle-Kreis, J. D. Surratt, S. Szidat, R. Szmigielski, and A. Wisthaler, "The Molecular Identification of Organic Compounds in the Atmosphere: State of the Art and Challenges," Chem. Rev. **115**, 3919–3983 (2015).
4. G. A. West, J. J. Barrett, D. R. Siebert, and K. V. Reddy, "Photoacoustic spectroscopy," Review of Scientific Instruments **54**, 797–817 (1983).
5. C. Haisch, "Photoacoustic spectroscopy for analytical measurements," Meas. Sci. Technol. **23**, 012001 (2011).
6. R. Lewicki, J. H. Doty, R. F. Curl, F. K. Tittel, and G. Wysocki, "Ultrasensitive detection of nitric oxide at 5.33 μm by using external cavity quantum cascade laser-based Faraday rotation spectroscopy," Proceedings of the National Academy of Sciences **106**, 12587–12592 (2009).
7. W. Zhao, G. Wysocki, W. Chen, E. Fertein, D. L. Coq, D. Petitprez, and W. Zhang, "Sensitive and selective detection of OH radicals using Faraday rotation spectroscopy at 2.8 μm," Opt. Express **19**, 2493–2501 (2011).
8. J. W. Daily, "Laser induced fluorescence spectroscopy in flames," Progress in Energy and Combustion Science **23**, 133–199 (1997).
9. A. A. Lanin, A. A. Voronin, A. B. Fedotov, and A. M. Zheltikov, "Time-domain spectroscopy in the mid-infrared," Sci Rep **4**, 6670 (2014).
10. A. A. Lanin, A. B. Fedotov, and A. M. Zheltikov, "Ultrabroadband XFROG of few-cycle mid-infrared pulses by four-wave mixing in a gas," J. Opt. Soc. Am. B **31**, 1901 (2014).
11. H. U. Stauffer, S. W. Grib, S. A. Schumaker, and S. Roy, "Broadband, background-free methane absorption in the mid-infrared," Opt. Express **29**, 21011 (2021).
12. I. Pupeza, M. Huber, M. Trubetskov, W. Schweinberger, S. A. Hussain, C. Hofer, K. Fritsch, M. Poetzlberger, L. Vamos, E. Fill, T. Amotchkina, K. V. Kepesidis, A. Apolonski, N. Karpowicz, V. Pervak, O. Pronin, F. Fleischmann, A. Azzeer, M. Žigman, and F. Krausz, "Field-resolved infrared spectroscopy of biological systems," Nature **577**, 52–59 (2020).
13. M. Liu, R. M. Gray, L. Costa, C. R. Markus, A. Roy, and A. Marandi, "Mid-infrared cross-comb spectroscopy," Nat Commun **14**, 1044 (2023).
14. T. Tomberg, A. Muraviev, Q. Ru, and K. L. Vodopyanov, "Background-free broadband absorption spectroscopy based on interferometric suppression with a sign-inverted waveform," Optica **6**, 147–151 (2019).
15. T. Buberl, P. Sulzer, A. Leitenstorfer, F. Krausz, and I. Pupeza, "Broadband interferometric subtraction of optical fields," Opt. Express **27**, 2432–2443 (2019).
16. W. Song, D. Okazaki, I. Morichika, and S. Ashihara, "Broadband background-free vibrational spectroscopy using a mode-locked Cr:ZnS laser," Opt. Express **30**, 38674 (2022).
17. T. T. Fricke, N. D. Smith-Lefebvre, R. Abbott, R. Adhikari, K. L. Dooley, M. Evans, P. Fritschel, V. V. Frolov, K. Kawabe, J. S. Kissel, B. J. J. Slagmolen, and S. J. Waldman, "DC readout experiment in Enhanced LIGO," Class. Quantum Grav. **29**, 065005 (2012).
18. T. F. Zehnpfennig, O. Shepherd, S. Rappaport, W. P. Reidy, and G. Vanasse, "Background suppression in double-beam interferometry," Appl. Opt. **18**, 1996–2002 (1979).

19. Z. Guan, M. Lewander, and S. Svanberg, "Quasi zero-background tunable diode laser absorption spectroscopy employing a balanced Michelson interferometer," Opt. Express **16**, 21714 (2008).
20. J. Hayden, S. Hugger, F. Fuchs, and B. Lendl, "A quantum cascade laser-based Mach–Zehnder interferometer for chemical sensing employing molecular absorption and dispersion," Appl. Phys. B **124**, 29 (2018).
21. I. Coddington, N. Newbury, and W. Swann, "Dual-comb spectroscopy," Optica **3**, 414–426 (2016).
22. A. S. Kowligy, H. Timmers, A. J. Lind, U. Elu, F. C. Cruz, P. G. Schunemann, J. Biegert, and S. A. Diddams, "Infrared electric field sampled frequency comb spectroscopy," Science Advances **5**, eaaw8794 (2019).
23. L. Ledezma, L. Ledezma, R. Sekine, Q. Guo, R. Nehra, S. Jahani, and A. Marandi, "Intense optical parametric amplification in dispersion-engineered nanophotonic lithium niobate waveguides," Optica **9**, 303–308 (2022).
24. M. Jankowski, M. Jankowski, M. Jankowski, N. Jornod, N. Jornod, C. Langrock, B. Desiatov, A. Marandi, M. Lončar, and M. M. Fejer, "Quasi-static optical parametric amplification," Optica **9**, 273–279 (2022).
25. M. Jankowski, A. Marandi, C. R. Phillips, R. Hamerly, K. A. Ingold, R. L. Byer, and M. M. Fejer, "Temporal Simultons in Optical Parametric Oscillators," Phys. Rev. Lett. **120**, 053904 (2018).
26. C. F. O'Donnell, S. C. Kumar, P. G. Schunemann, and M. Ebrahim-Zadeh, "Femtosecond optical parametric oscillator continuously tunable across 3.6-8 µm based on orientation-patterned gallium phosphide," Opt. Lett. **44**, 4570–4573 (2019).
27. M. Liu, R. M. Gray, A. Roy, K. A. Ingold, E. Sorokin, I. Sorokina, P. G. Schunemann, and A. Marandi, "High-Power Mid-IR Few-Cycle Frequency Comb from Quadratic Solitons in an Optical Parametric Oscillator," Laser & Photonics Reviews 2200453 (2022).
28. Q. Ru, T. Kawamori, P. G. Schunemann, S. Vasilyev, S. B. Mirov, S. B. Mirov, and K. L. Vodopyanov, "Two-octave-wide (3–12 µm) subharmonic produced in a minimally dispersive optical parametric oscillator cavity," Opt. Lett. **46**, 709–712 (2021).
29. J. Zhang, C. Ning, J. Heng, S. Yu, and Z. Zhang, "Ultra-Short Pulse Generation From Optical Parametric Oscillators With a Cavity-Length Detuning," IEEE Photonics Technology Letters **34**, 263–266 (2022).
30. A. Marandi, N. C. Leindecker, V. Pervak, R. L. Byer, and K. L. Vodopyanov, "Coherence properties of a broadband femtosecond mid-IR optical parametric oscillator operating at degeneracy," Opt. Express **20**, 7255–7262 (2012).
31. H. Harde, S. Keiding, and D. Grischkowsky, "THz commensurate echoes: Periodic rephasing of molecular transitions in free-induction decay," Phys. Rev. Lett. **66**, 1834–1837 (1991).
32. I. Coddington, W. C. Swann, and N. R. Newbury, "Time-domain spectroscopy of molecular free-induction decay in the infrared," Opt. Lett. **35**, 1395–1397 (2010).
33. N. R. Newbury, I. Coddington, and W. Swann, "Sensitivity of coherent dual-comb spectroscopy," Opt. Express **18**, 7929–7945 (2010).
34. R. Hamerly, A. Marandi, M. Jankowski, M. M. Fejer, Y. Yamamoto, and H. Mabuchi, "Reduced models and design principles for half-harmonic generation in synchronously pumped optical parametric oscillators," Phys. Rev. A **94**, 063809 (2016).
35. I. E. Gordon, L. S. Rothman, R. J. Hargreaves, R. Hashemi, E. V. Karlovets, F. M. Skinner, E. K. Conway, C. Hill, R. V. Kochanov, Y. Tan, P. Wcisło, A. A. Finenko, K. Nelson, P. F. Bernath, M. Birk, V. Boudon, A. Campargue, K. V. Chance, A. Coustenis, B. J. Drouin, J. –M. Flaud, R. R. Gamache, J. T. Hodges, D. Jacquemart, E. J. Mlawer, A. V. Nikitin, V. I. Perevalov, M. Rotger, J. Tennyson, G. C. Toon, H. Tran, V. G. Tyuterev, E. M. Adkins, A. Baker, A. Barbe, E. Canè, A. G. Császár, A. Dudaryonok, O. Egorov, A. J. Fleisher, H. Fleurbaey, A. Foltynowicz, T. Furtenbacher, J. J. Harrison, J. –M. Hartmann, V. –M. Horneman, X. Huang, T. Karman, J. Karns, S. Kassi, I. Kleiner, V. Kofman, F. Kwabia–Tchana, N. N. Lavrentieva, T. J. Lee, D. A. Long, A. A. Lukashevskaya, O. M. Lyulin, V. Yu. Makhnev, W. Matt, S. T. Massie, M. Melosso, S. N. Mikhailenko, D. Mondelain, H. S. P. Müller, O. V. Naumenko, A.

Perrin, O. L. Polyansky, E. Raddaoui, P. L. Raston, Z. D. Reed, M. Rey, C. Richard, R. Tóbiás, I. Sadiek, D. W. Schwenke, E. Starikova, K. Sung, F. Tamassia, S. A. Tashkun, J. Vander Auwera, I. A. Vasilenko, A. A. Vigasin, G. L. Villanueva, B. Vispoel, G. Wagner, A. Yachmenev, and S. N. Yurchenko, "The HITRAN2020 molecular spectroscopic database," Journal of Quantitative Spectroscopy and Radiative Transfer **277**, 107949 (2022).

36. J. Zhang, A. P. Shreenath, M. Kimmel, E. Zeek, R. Trebino, and S. Link, "Measurement of the intensity and phase of attojoule femtosecond light pulses using Optical-Parametric-Amplification Cross-Correlation Frequency-Resolved Optical Gating," Opt. Express **11**, 601–609 (2003).
37. P. M. Vaughan and R. Trebino, "Optical-parametric-amplification imaging of complex objects," Opt. Express **19**, 8920–8929 (2011).
38. E. Mimoun, L. De Sarlo, J.-J. Zondy, J. Dalibard, and F. Gerbier, "Sum-frequency generation of 589 nm light with near-unit efficiency," Opt. Express **16**, 18684 (2008).
39. A. Roy, L. Ledezma, L. Costa, R. Gray, R. Sekine, Q. Guo, M. Liu, R. M. Briggs, and A. Marandi, "Visible-to-mid-IR tunable frequency comb in nanophotonics," Nat Commun **14**, 6549 (2023).

**Acknowledgements**

The authors gratefully acknowledge support from NSF grant no. 1846273, AFOSR award FA9550-20-1-0040, the center for sensing to intelligence at Caltech, and NASA/JPL.

**Competing interests**

AM and LL have financial interest in PINC Technologies Inc., which is developing photonic integrated nonlinear circuits.